\documentclass[12pt]{iopart}

\usepackage{graphicx}
\begin{document}

\title[The role of the Rashba coupling in spin current of monolayer gapped graphene.]{The role of
the Rashba coupling in spin current of monolayer gapped graphene.}

\author{K. Hasanirokh, J. Azizi, A. Phirouznia, H. Mohammadpour}

\address{Department of Physics, Azarbaijan Shahid Madani University, 53714-161, Tabriz, Iran}
\begin{abstract}
In the current work we have investigated the influence of the Rashba
spin-orbit coupling on spin-current of a single layer gapped
graphene. It was shown that the Rashba coupling has a considerable
role in generation of the spin-current of vertical spins in
mono-layer graphene. The behavior of the spin-current is determined
by density of impurities. It was also shown that the spin-current of
the system could increase by increasing the Rashba coupling strength
and band-gap of the graphene and the sign of the spin-current could
be controlled by the direction of the current-driving electric
field.
\end{abstract}

\maketitle

\section{Introduction}
Carbon nano materials reveal an interesting polymorphism of
different allotropes exhibiting each possible dimensionality: 1)
fullerene molecule (0D), 2) nano tubes (1D),  3) graphene and
graphite platelets (2D) and 4) diamond (3D) are selected
examples.
\\
Since graphene was discovered in 2004 by Geim and his team
\cite{Novoselov2}, it has continued to surprise scientists. This is
due to some spectacular and fantastic transport properties like the
high carrier mobility and universal minimal conductivity at the
Dirac point where the valence and conduction bands cross each other
\cite{Novoselov1,Tombros}.
\\
In the presence of spin-orbit couplings, spin polarized states in
graphene is a response to an external in-plane electron field. In
today's spintronics, it is a possible key ingredient towards
electrical spin control with spin-orbit interaction which has
attracted a great attention, such as anomalous Hall effect
\cite{Han}. Extrinsic spin-orbit interaction (Rashba) that
originates from the structure inversion asymmetry (SIA), arises from
symmetry breaking generated at interface between graphene and
substrate or by external fields \cite{Rashba,Rashba1,Huertas}.
\\
Spin-current control is of crucial importance in electronic devices.
Some attempts have been made for applying Rashba interaction to
control spin-current. Since Rashba coupling strength in graphene is
high relative to other materials, studying the role of this
interaction in graphene could remove most of the existing obstacles
in spin-current control. As mentioned in previous studies, amount of
Rashba interaction in graphene can be at a very high. For example,
0.2 ev which has been reported in \cite{Dedkov} can be pointed out. \\
Depending on graphene's substrate, strength of the Rashba
interaction in graphene can be much higher than its intrinsic
spin-orbit interaction. In addition to the spin transport related
aspects generated by applying the Rashba interaction, this coupling
can be utilized in optical devices as well. As an example, by
applying this interaction in two-dimensional electron gas and
graphene, controllable blue shift could be obtained in absorption
spectrum \cite{Phirouznia,Jamshidi}. Spin-current is an important
tool for studying spin related features in graphene and is also
important in the development of the graphene quantum computer.
\\
Based on the Tight-Binding model, Soodchomshom has studied the
effect of the uniaxial strain in the spin transport through a
magnetic barrier of the strained graphene system and shown that
graphene has a fantastic potential for applications in
nano-mechanical spintronic devices. Strain in graphene will induce
the pseudo-potentials at the barrier that can control the spin
currents of the junction \cite{Bumned}. Ezawa has considered a nano
disk connected with two leads and shown this system acts as a spin
filter and can generate a spin-current \cite{Motohiko}.
\\
In this work, we have considered the influence of Rashba spin-orbit
coupling on spin-current of a monolayer gapped graphene. It has been
found that the spin-current associated with normal spins is
influenced by the Rashba spin-orbit coupling and the energy gap of
the system.
\section{Model and approach}
The low energy charge carriers in graphene  are satisfied in a
massless Dirac equation. This equation have an isotropic linear
energy dispersion near the Dirac points \cite{Novoselov2}. Meanwhile
the Hamiltonian of gapped graphene with Rashba spin-orbit (SO)
coupling is \cite{Giavaras,Kane}
\begin{equation}
\hat{H}=\hat{H}_{G}+\hat{H}_{R} + \hat{H}_{gap}+\hat{V}_{im}.
\end{equation}
$\hat{H}_{G}$  is the Dirac Hamiltonian for massless fermions which
is given as follows \cite{Ahmadi,Rashba3}
\begin{equation}
 \hat{H}_{G}=-i\gamma\psi^{\dag}(\sigma_x\tau_z\partial x+\sigma_y\partial y)\psi,
\end{equation}
$v_f$ is the Fermi velocity in graphene, $\gamma=\hbar v_f$,
$\sigma$ and $\tau$ represent the Pauli matrices where,
$\sigma_z=\pm1,\tau_z=\pm1$ denote the Pauli matrice of pseudospin
on the A and B sublattices and different Dirac points, respectively.
\\
The second term in equation (1), $H_R$, is the Rashba (SO)
interaction, and can be expressed as \cite{Kane,Yi}
\begin{equation}
 \hat{H}_{R}=\frac{\lambda_R}{2}\psi^{\dag}(\sigma_y s_x-\sigma_x\tau_zs_y)\psi,
\end{equation}
where $\lambda_R$ is the (SO) coupling constant and $\textbf{s}$
denotes Pauli matrix representing the spin of electron. In an ideal
graphene sheet, the Dirac electrons are massless and the band
structure has no energy gap. Experimentally, it is possible to
manipulate an energy gap (from a few to hundreds of meV) in
graphene's band structure, namely a Dirac gap
\cite{Vitali,Zhou,Enderlein}. As a result of the asymmetry in
graphene sublattices, A and B, the band gap can have a nonzero
value. The last term in $H_{gap}=\tau\Delta\sigma_z$ equation (1),
referred to mass term, arises from the energy gap $\Delta$ in the
spectrum of graphene, that $\tau =1$ $(\tau=-1)$ corresponds to the
$K$ ($K'$) valley. The last term in equation (1), $\hat{V}_{im}$, is
induced due to the short range impurities which can be written as
\begin{eqnarray}
 V_{im}(r)=\sum_jV \delta({\vec{r}-\vec{r}_j}),
\end{eqnarray}
where the summation is over the position of impurities. The
eigenfunctions of $ \hat{H}_0=\hat{H}_{G}+\hat{H}_{R} +
\hat{H}_{gap}$ are
\begin{eqnarray}
\mid \epsilon_1(k)>=\left(
\begin{array}{cccc}
-i\Omega_1(k)e^{-2i\varphi}\\
\Omega_2(k)e^{-i\varphi}\\
i\Omega_3(k)e^{-i\varphi}\\
1
\end{array}
\right),
\end{eqnarray}
\begin{eqnarray}
\mid \epsilon_2(k)>=\left(
\begin{array}{cccc}
i\Omega_4(k)e^{-2i\varphi}\\
\Omega_5(k)e^{-i\varphi}\\
i\Omega_3(k)e^{-i\varphi}\\
1
\end{array}
\right),
\end{eqnarray}
\begin{eqnarray}
\mid \epsilon_3(k)>=\left(
\begin{array}{cccc}
i\Omega_6(k)e^{-2i\varphi}\\
\Omega_7(k)e^{-i\varphi}\\
-i\Omega_8(k)e^{-i\varphi}\\
1
\end{array}
\right),
\end{eqnarray}
\begin{eqnarray}
\mid \epsilon_4(k)>=\left(
\begin{array}{cccc}
-i\Omega_9(k)e^{-2i\varphi}\\
\Omega_{10}(k)e^{-i\varphi}\\
-i\Omega_8(k)e^{-i\varphi}\\
1
\end{array}
\right),
\end{eqnarray}
\\
and the corresponding eigenvalues are
\\
\begin{eqnarray}
\epsilon_1(k)=-\sqrt{\Delta^2+\gamma^2
k^2+\frac{\lambda^2}{8}-\frac{\lambda}{8}\Gamma(k)}\nonumber\\
\epsilon_2(k)=-\epsilon_1(k)\\
\epsilon_3(k)=-\sqrt{\Delta^2+\gamma^2
k^2+\frac{\lambda^2}{8}+\frac{\lambda}{8}\Gamma(k)}\nonumber\\
\epsilon_4(k)=-\epsilon_3(k)\nonumber
\end{eqnarray}
where we have defined
\begin{eqnarray}
\Omega_1(k)=-\Lambda_+(k)\xi^{(1)}_+(k) ~~~~~
\Omega_2(k)=\xi^{(1)}_+(k)\nonumber\\
\Omega_3(k)=\Lambda_-(k)~~~~~~~~~~~~~~~~
\Omega_4(k)=\Lambda_+(k)\xi^{(1)}_-(k)\\
\Omega_5(k)=\xi^{(1)}_-(k)~~~~~~~~~~~~~~~~ \Omega_6(k)=-\Lambda_-(k)
\xi^{(2)}_+(k)\nonumber\\
\Omega_7(k)=\xi^{(2)}_+(k)~~~~~~~~~~~~~~~~
\Omega_8(k)=\Lambda_+(k) \nonumber\\
\Omega_9(k)=\Lambda_-(k)\xi^{(2)}_-(k)~~~~~~~~
\Omega_{10}(k)=\xi^{(2)}_-(k)\nonumber
\end{eqnarray}
in which  $\Gamma(k)=(16\gamma^2 k^2+\lambda^2)^{1/2}$,
$\Lambda_{\pm}(k)=(\pm\lambda+\Gamma(k))/(4\gamma k)$,
$\xi^{(1)}_{\pm}(k)=(\Delta\pm \epsilon_1(k))/(\gamma k)$ and
$\xi^{(2)}_{\pm}(k)=(\Delta\pm \epsilon_3(k))/(\gamma k)$.
\\
The transition probabilities between the $|\epsilon_i(k)>$ and
$|\epsilon_j(k')>$ states are given by the Fermi's golden rule,
\begin{equation}
\omega_{ij}(\vec{k},\vec{k'})=(2 \pi /\hbar)
n_{i}|<\vec{k'_j}|\hat{V_{im}}|\vec{k_i}>|^2
\delta(\epsilon_{i}(\vec{k})-\epsilon_{j}(\vec{k'})),~~~~~(i,j=1,2,3,4).
\end{equation}
that
$\delta(\epsilon_{i}(\vec{k})-\epsilon_{j}(\vec{k'}))\approx\delta({\vec{k}}-{\vec{k'}})/\gamma$.
\\
Scattering potential of the impurities is described by the
$\hat{V_{im}}$. These are assumed to be distributed randomly with
real density $n_{i}$
\begin{equation}
\bar{\omega}_i=K\int d\acute{\varphi}\sum_j
\omega_{ji}(\varphi,\varphi'),~~~~~(i,j=1,2,3,4)
\end{equation}
in which $K=(2\pi/\hbar)n_{i}$ and $\varphi$, $\varphi'$ are two
angles characterizing the direction of $\vec{k}$ and $\vec{k'}$
states relative to the $x$ axis respectively. Then the
non-equilibrium distribution function of a given $\epsilon_i(k)$
energy band can be written as \cite{Karel}
\begin{equation}
\delta f_i=-ev_iE(-\partial_\epsilon f_0)[a_i(\varphi)
cos\theta+b_i(\varphi) sin\theta],
\end{equation}
in which $E$ is the current driven electric field, $\theta$ is the
angle between the $x$ axis and the direction of the electric field
and the unknown functions $a_i(\varphi)$ and $b_i(\varphi)$ can be
written as follows
\begin{eqnarray}
a_i(\varphi)=a_0+\sum_{j=1} a_{cji}\cos(j\varphi)+\sum_{j=1}
a_{sji}\sin(j\varphi)\\
b_i(\varphi)=b_0+\sum_{j=1} b_{cji}\cos(j\varphi)+\sum_{j=1}
b_{sji}\sin(j\varphi)
\end{eqnarray}
$a_{i}(\varphi)$ and $b_{i}(\varphi)~~~(i=1,2,3,4)$ must satisfy
\cite{Karel}
\begin{equation}
\label{main1} \cos\varphi=\bar{\omega}_i a_{i}(\varphi)-K\int
d\acute{\varphi}\sum_j [\omega_{ij}(\varphi,\varphi')a_{j}(\varphi)]
\end{equation}
\begin{equation}
\label{main2} \sin\varphi=\bar{\omega}_i b_{i}(\varphi)-K\int
d\acute{\varphi}\sum_j [\omega_{ij}(\varphi,\varphi')b_{j}(\varphi)]
\end{equation}
it can be easily obtained that the non-vanishing parameters are
$a_{c1i}$ and $b_{s1i}$ in which $a_{c_{1i}}=b_{s_{1i}} ~~(i=1-4).$
This can be inferred from the Dirac point approximation in which at
this level band energies are appeared to be isotropic in k-space.
Meanwhile, these parameters can be obtained through the equations
\eref{main1}-\eref{main2} The non-equilibrium distribution function
is then given by
\begin{equation}
\label{f}
 \delta f_i=-ev_iE(-\partial_\epsilon f_0)[a_{c_{1i}}\cos\varphi
cos\theta+b_{s_{1i}} \sin\varphi sin\theta]
\end{equation}
The spin current operator is defined
as \cite{Rashba}
\begin{equation}
J^{s_i}_{j}=\{s_i,\hat{v}_j\},
\end{equation}
where $\hat{v}_j=\hbar^{-1}(\frac{\partial \hat{H}}{\partial k_j})
~~(i,j=x,y)$ is the velocity operator. The spin current operator in
the basis $e^{(i\vec{k}.\vec{r})}|s\sigma>$ is given as follows,
\begin{equation}
J^{s_z}_{x}=\left(
\begin{array}{cccc}
0&\hbar\gamma&0&0\\
\hbar\gamma&0&0&0\\
0&0&0&-\hbar\gamma\\
0&0&\hbar\gamma&0
\end{array}
\right),
\end{equation}
\begin{equation}
J^{s_z}_{y}=\left(
\begin{array}{cccc}
0&-i\hbar\gamma&0&0\\
i\hbar\gamma&0&0&0\\
0&0&0&i\hbar\gamma\\
0&0&-i\hbar\gamma&0
\end{array}
\right).
\end{equation}
Then the average spin currents in
x and y directions are given by
\begin{equation}
\label{J} <J^{s_i}_{j}>=\frac{1}{(2\pi)^2}\int
d^2k\sum_{\lambda=1}^4<k\lambda\mid J^{s_i}_{j}\mid k\lambda >
f_{\lambda}(k) ~~~~(i,j=x,y),
\end{equation}
using the equations \eref{f} and \eref{J}, one can obtain
\begin{eqnarray}
<J^{s_z}_{x}>=&-&J^s_0(\frac{\epsilon_2(k_f)b_{s12}k_f}
{\gamma-\frac{\gamma\lambda}{\Gamma(k_f)}}[\Omega_4(k_f)\Omega_5(k_f)-\Omega_3(k_f)][\Omega_3(k_f)\Omega_4(k_f)+\Omega_5(k_f)]\nonumber\\
&+&\frac{\epsilon_4(k_f)b_{s14}k_f}{\gamma+\frac{\gamma\lambda}{\Gamma(k_f)}}[\Omega_8(k_f)-\Omega_9(k_f)
\Omega_{10}(k_f)][\Omega_8(k_f)\Omega_9(k_f)+\Omega_{10}(k_f)]),\nonumber\\
\end{eqnarray}
where $J^s_0=\frac{\hbar eE}{\pi}$.
\\
Simil1ary one can easily obtain other components of the spin-current as follows\\
\begin{eqnarray}
<J^{s_z}_{y}>=-<J^{s_z}_{x}>,\nonumber\\
<J^{s_x}_{x}>=<J^{s_x}_{y}>=0,\\
<J^{s_y}_{x}>=<J^{s_y}_{y}>=0. \nonumber
\end{eqnarray}
This means that the Rashba coupling cannot generate spin current of
in-plane spin components ($J^{s_x}_i$ and $J^{s_y}_i$). This is in
agreement with the results of the similar case in two-dimensional
electron gas in which the Rashba coupling induced spin current
identically vanishes \cite{Zhian,Inoue}.
\\
 Electrical current will be as following
\begin{eqnarray}
<J_{x}>&=&\frac{1}{(2\pi)^2}\int d^2k\sum_{\lambda=1}^4<k\lambda\mid e\hat{v}_x\mid k\lambda > f_{\lambda}(k)=\nonumber\\
\nonumber\\
&-&J_0(\frac{\epsilon_2(k_f)a_{c12}k_f}{\gamma-\frac{\gamma\lambda}
{\Gamma(k_f)}}[\Omega_3(k_f)\Omega_4(k_f)+\Omega_5(k_f)]^2+\nonumber\\
\nonumber\\
&&\frac{\epsilon_4(k_f)a_{c14}k_f}{\gamma+\frac{\gamma\lambda}
{\Gamma(k_f)}} [\Omega_8(k_f)\Omega_9(k_f)+\Omega_{10}(k_f)]^2),
\end{eqnarray}
that $J_0=(\hbar e^2Ek_f)/(\pi \gamma).$\\
Calculation will obtain the following result, directly
$<J_{x}>=<J_{y}>$ which can be regarded as a consequence of the
Dirac point approximation that could eliminate the anisotropic
effects of the band structure such as trigonal warping.

\section{Results}
Here, the spin-current of a monolayer gapped graphene has been
obtained in the presence of Rashba interaction. In this work, it was
shown that the non-equilibrium spin-current of vertical spins can be
effectively controlled by this spin-orbit interaction.
\\
It was assumed that the electrical field has been applied along the
$x$ axis and the numerical parameters have been chosen as follows
$\epsilon_f=1meV$ is the Fermi energy \cite{Liewrian} and
$n_i=10^{10}cm^{-2}$ is the density of impurities.
\\
Different non-equilibrium spin-current components have been depicted
as a function of the graphene gap in figures 1-2. These figures
clearly show that the longitudinal and transverse non-equilibrium
spin-currents of normal spins have accountable values in which their
signs and magnitudes can be controlled by the graphene's gap. The
absolute value of non-equilibrium spin-current components, with
respect to the gap of graphene, are increasing at by increasing the
Rashba coupling strength. One of the important features which can be
inferred from the figures 1 and 2 is the fact that the spin current
can be of either sign, depending on the direction of the driving
electric field.
\\
\begin{figure}
\includegraphics[width=4.0in]{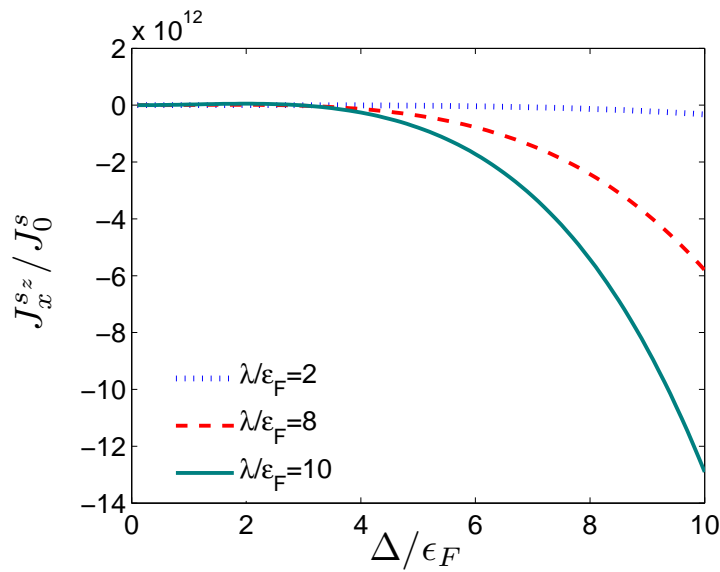}
\caption{non-equilibrium longitudinal spin current as a function of
the graphene gap at different Rashba couplings.}
 \label{fig1}
\end{figure}
\\
\begin{figure}
\includegraphics[width=4.0in]{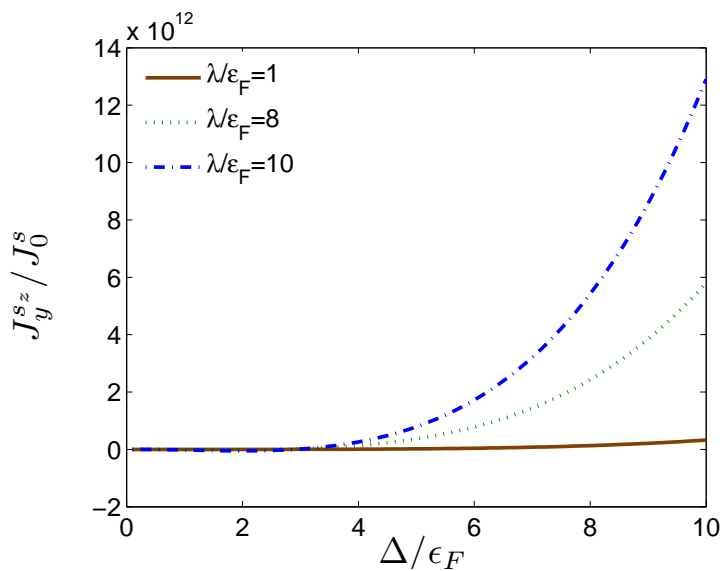}
\caption{non-equilibrium transverse spin current as a function of
the graphene gap at different Rashba couplings.}
 \label{fig2}
\end{figure}
Figure 3 displays the electric current along the x direction as a
function of the gap. As illustrated in this figure the electric
current in gaped graphene can be effectively changed by the Rashba
coupling. The absolute value of  electrical current increases by
increasing the amount of the gap. The deference between the curves
inside this figure demonstrates the importance of the Rashba
coupling.
\begin{figure}
\includegraphics[width=4.0in]{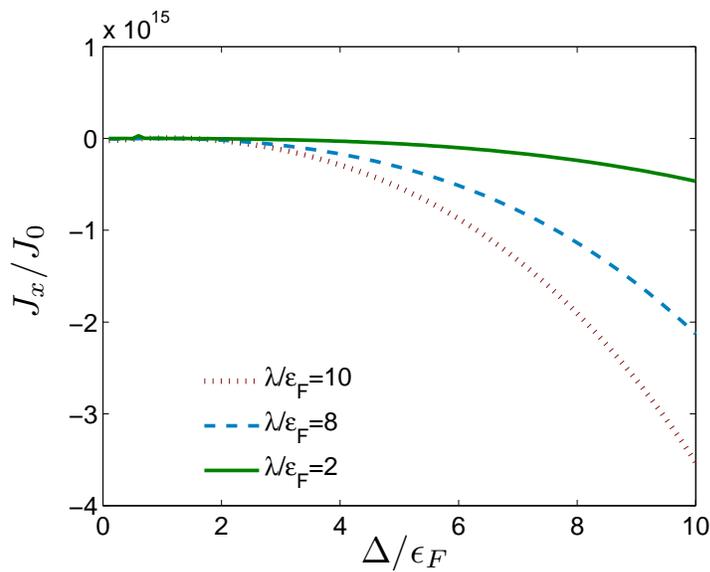}
\caption{Longitudinal electric current as a function of the gap in
graphene at different Rashba couplings.}
 \label{fig3}
\end{figure}
The longitudinal non-equilibrium spin-currents of a monolayer gapped
graphene have indicated as a function of the Rashba coupling in
figure 4. The Rashba spin-orbit coupling strength can reach high
values up to 0.2eV in monolayer graphene. As shown in this figure,
the absolute value of spin-current induced by the Rashba coupling
increases by increasing the gap.
\\
As can be seen in figures 1-4,  absolute value of spin-current would
increase by increasing the Rashba coupling strength and also the
energy gap because, on the one hand, the increased energy gap would
reduce the possibility of spin relaxation and spin mixing and on the
other hand, increased Rashba coupling strength would raise effective
magnetic field of this coupling. This effective magnetic field can
be regarded as
$\textbf{B}_{eff}=\lambda_R/(2\mu_B)(\sigma_y\hat{x}-\sigma_x\hat{y})$.
Anisotropy induced by current-driving electric field results in a
non-vanishing average effective magnetic field; i.e. if the
current-driving electric field is along with $x$, hopping along with
$x$ axis would be more likely to happen and
$<\sigma_x>~>~<\sigma_y>$ therefore the existing electrons at Fermi
level would feel a non-zero effective magnetic field where the
spin-current is originating from this field. Therefore, external
in-plane electric field plays an important role in generating the
effective magnetic field on spin carriers. Consequently, it is
expected that, if bias voltage is applied along the $y$ axis, the
direction of effective magnetic field would also change; thus, type
of spin majority carriers would also modify. This phenomenon can be
clearly seen in figures 1-4 so that spin-current sign changes
depending on the direction of the applied bias voltage. Therefore,
the sign of spin current would be controllable by external bias.
According to the mentioned points, generating spin-current of
vertical spins at least in non-equilibrium regime can be expected.
\\
The behavior of the spin-current is determined by the impurity
density as depicted in figure 5. In this figure, we have taken
$\Delta/\epsilon_F=5$ and it can be inferred from the data depicted
in this figure that increasing the spin-mixing rate (which could
take place by increasing the density of impurities) decreases the
spin-polarization and spin-current of the system.
\begin{figure}
\includegraphics[width=4.0in]{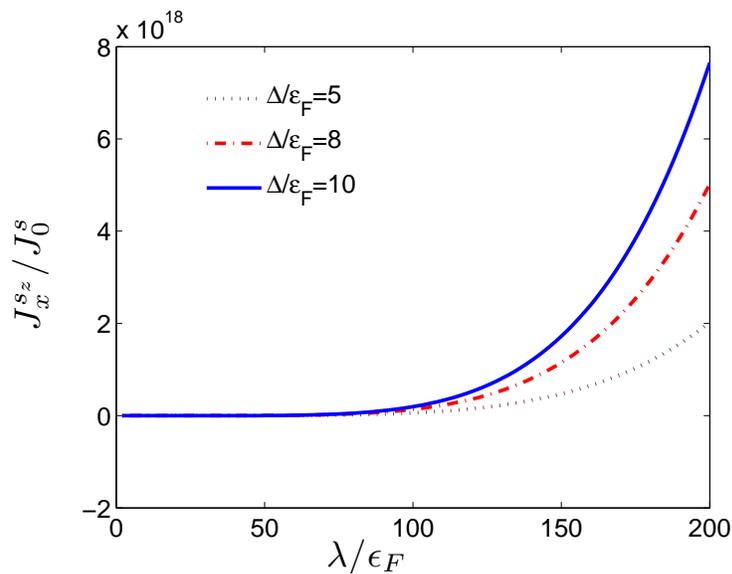}
\caption{Longitudinal spin current as a function of the Rashba coupling.}
 \label{fig4}
\end{figure}
\begin{figure}
\includegraphics[width=4.0in]{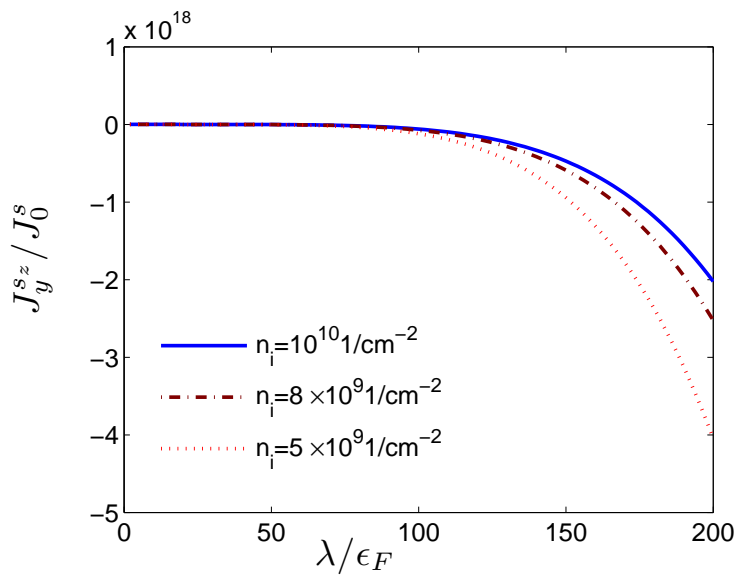}
\caption{Transverse spin current as a function of the Rashba
coupling at $\Delta/\epsilon_F=5$.}
 \label{fig5}
\end{figure}
\\
\section{Conclusion}
In the present work, the influence of the Rashba coupling on
spin-related transport effects have been studied. Results of the
present study show that the Rashba interaction has an important role
in generation of the non-equilibrium spin-current of vertical spins
in a monolayer gapped graphene. The absolute value of spin-current
as a function of the gap, increases by increasing the Rashba
interaction strength in non-equilibrium regime. Another important
point in the results of the present study can be describe as
follows; Not only the amount of spin-current in grapheme is
controllable by gate voltage (responsible for Rashba interaction)
but also its sign is predictable by the direction of the applied
 bias voltage.
\section*{References}
\bibliographystyle{model1a-num-names}
\bibliography{apssamp}

\end{document}